\documentclass[prl,twocolumn,showpacs,floats]{revtex4}
\usepackage{mathrsfs}

\usepackage{graphicx}

\begin{document}

\title{Electronic Evidence of Unusual Magnetic Ordering in a Parent Compound of FeAs-Based Superconductors }

\author{Guodong  Liu$^{1}$, Haiyun Liu$^{1}$,  Lin Zhao$^{1}$, Wentao Zhang$^{1}$,  Xiaowen Jia$^{1}$,
Jianqiao Meng$^{1}$, Xiaoli Dong$^{1}$, G. F. Chen$^{1}$, Guiling
Wang$^{2}$, Yong Zhou$^{2}$, Yong Zhu$^{2}$, Xiaoyang Wang$^{2}$,
Zuyan Xu$^{2}$, Chuangtian Chen$^{2}$ and X. J. Zhou$^{1,*}$}

\affiliation{
\\$^{1}$Beijing National Laboratory for Condensed Matter Physics, Institute of Physics,
Chinese Academy of Sciences, Beijing 100190, China
\\$^{2}$Technical Institute of Physics and Chemistry, Chinese Academy of Sciences, Beijing 100190, China
}
\date{March 25, 2009}
%
%

\begin{abstract}

High resolution angle-resolved photoemission measurements have been
carried out on BaFe$_2$As$_2$, a parent compound of the FeAs-based
superconductors.  In the magnetic ordering state, there is no gap
opening observed on the Fermi surface. Instead, dramatic band
structure reorganization occurs across the magnetic transition. The
appearance of the singular Fermi spots near ($\pi$,$\pi$) is the
most prominent signature of magnetic ordering. These observations
provide direct evidence that the magnetic ordering state of
BaFe$_2$As$_2$ is distinct from the conventional spin-density-wave
state.  They reflect the electronic complexity in this
multiple-orbital system and necessity in involving the local
magnetic moment in describing the underlying electron structure.

\end{abstract}

\pacs{74.70.-b, 74.25.Jb, 79.60.-i, 71.20.-b}

\maketitle

The interplay between magnetism and superconductivity is a prominent
topic in the newly-discovered FeAs-based
superconductors\cite{Kamihara,ZARenSm,RotterSC}. The ground state of
the parent compound of the FeAs-based superconductors has a
collinear anti-ferromagnetic
ordering\cite{DongSDW,PCDai,BFSNeutron3}. Superconductivity is
achieved from suppressing the magnetic ordering by chemical
doping\cite{Kamihara,ZARenSm,RotterSC} or application of high
pressure\cite{PressureTc}. Magnetism has been proposed to play an
important role in generating high temperature superconductivity in
this new class of high temperature
superconductors\cite{FeSCMagnetic,Triplet}. However, the nature of
the magnetic ordering and its origin remain under hot
debate\cite{DongSDW,J1J2Picture}. It is proposed that the magnetic
ordering is associated with the spin-density-wave (SDW) formation
driven by the Fermi surface nesting\cite{DongSDW}. Recent
optical\cite{WZHuOpt} and quantum oscillation\cite{Oscillations}
measurements reported gap opening related with the magnetic
transition which is in favor of the SDW picture. It becomes critical
to pin down whether there is really a gap opening and, if there is,
where the gap opens in the momentum space. In principle,
angle-resolved photoemission (ARPES) measurements can provide direct
information on these issues but so far the ARPES results remain
inconclusive\cite{DLFBFe2As2,CLiuBFeAs,HYLiu,LZhao,HDing,Borisenko,DLFSFA,ZHasanSFS,MYi}.

In this paper we report angle-resolved photoemission evidence to
show the unusual nature of magnetic ordering in a parent compound of
FeAs-based superconductors, BaFe$_2$As$_2$\cite{RotterParent}.
Unlike the conventional SDW system\cite{CrSDW} where magnetic
ordering is associated with Fermi surface nesting and a gap
opening\cite{CrARPES}, the electronic structure of BaFe$_2$As$_2$
shows no sign of gap opening on the Fermi surface in the
magnetically-ordered state. Instead, it shows a dramatic band
reorganization and particularly an emergence of singular strong
Fermi spots. These observations provide direct and clear-cut
evidences that the magnetic ordering state of BaFe$_2$As$_2$ is
distinct from the conventional SDW state.  They also point to the
importance and necessity in involving the local magnetic moment in
the underlying electron structure.  These results and their
understanding will have important implications on understanding the
magnetism and superconductivity in this new class of
superconductors.

The ARPES measurements are carried out on our photoemission system
equipped with Scienta R4000 electron energy analyzer and Helium
discharge lamp which gives a photon energy of h$\upsilon$= 21.218
eV\cite{GDLiu}. The light on the sample is partially polarized. The
energy resolution was set at 10 meV and the angular resolution is
$\sim$0.3 degree. The Fermi level is referenced by measuring on the
Fermi edge of a clean polycrystalline gold that is electrically
connected to the sample. The BaFe$_2$As$_2$ single crystals were
grown using flux method\cite{WZHuOpt} which show a
magnetic$\slash$structural transition at T$_{MS}$$\sim$138
K\cite{RotterParent,WZHuOpt}.  The crystals were cleaved {\it in
situ} and measured in vacuum with a base pressure better than
5$\times$10$^{-11}$ Torr.  For ARPES measurements, we found that the
cleaved surface of BaFe$_2$As$_2$ shows a change with time. To
reduce the influence of this aging effect, we took many independent
measurements on fresh surfaces. All the Fermi surface mapping
measurements reported here were taken within four hours after
cleaving (Fig. 1), and for the typical momentum cuts (Fig. 2), the
measurements were carried out within two hours after cleaving.

\begin{figure}[tbp]
\begin{center}
\includegraphics[width=0.95\columnwidth,angle=0]{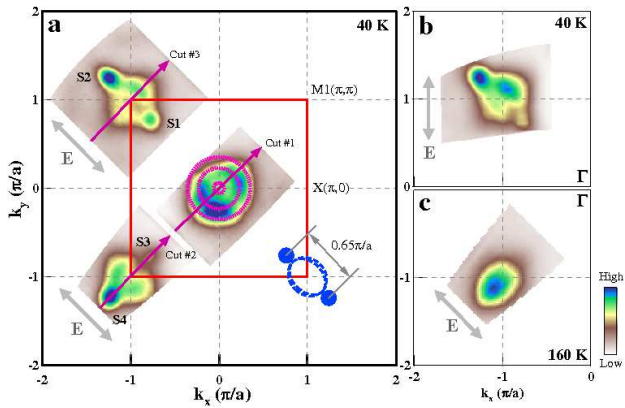}
\end{center}
\caption{Fermi surface of BaFe$_2$As$_2$ above and below the
magnetic transition temperature (T$_{MS}$=138 K).  (a).Spectral
weight distribution integrated within [-10meV,10meV] energy window
with respect to the Fermi level as a function of k$_x$ and k$_y$
measured at 40 K.  All the data were taken with the major electric
vector along the ($\pi$,-$\pi$)$-$(-$\pi$,$\pi$) diagonal direction,
as shown by the double-headed arrows. (b). Fermi surface near the M2
point measured at 40 K with the main electric vector along
(0,0)-(0,$\pi$) vertical direction. (c). Fermi surface near
M3(-$\pi$,-$\pi$) measured at 160 K. }
\end{figure}

In the magnetic ordering state,  the Fermi surface of
BaFe$_{2}$As$_{2}$ near the $\Gamma$ point shows three hole-like
sheets (Fig. 1a), as evidenced from Fig. 2a and Fig. 2d where three
Fermi crossings are clearly resolved in the bands taken along the
$\Gamma$$-$M1($\pi$,$\pi$) direction (Cut $\#$1 in Fig. 1a. For the
convenience of description, we denote the four M points as
M1($\pi$,$\pi$), M2(-$\pi$,$\pi$), M3(-$\pi$,-$\pi$) and
M4($\pi$,-$\pi$)). The corresponding Fermi momenta are $\sim$0.06,
$\sim$0.20 and $\sim$0.35$\pi$/a, respectively. Additional
measurements along $\Gamma$$-$X($\pi$,0) and other directions also
give three Fermi crossings with similar Fermi momenta.

\begin{figure}[tbp]
\begin{center}
\includegraphics[width=0.95\columnwidth,angle=0]{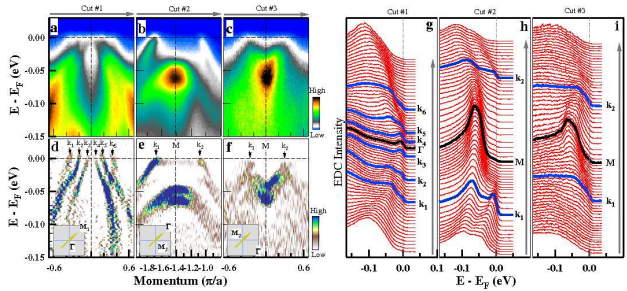}
\end{center}
\caption{Band structure and photoemission spectra of BaFe$_2$As$_2$
along typical high-symmetry cuts measured at 20 K. (a), (b) and (c)
show original photoemission images measured along the three high
symmetry cuts (Cuts $\#$1, 2 and 3 in Fig. 1a). The location of the
three cuts are also shown in the insets of Fig.2(d-f). To highlight
underlying band structure more clearly, (d) (e) and (f) show
corresponding second derivative images. (g), (h) and (i) show
photoemission spectra (energy distribution curves, EDCs) for the
three cuts with those near the Fermi momenta and high symmetry
points colored and marked. }
\end{figure}

The most prominent features in the Fermi surface topology of
BaFe$_2$As$_2$ below T$_{MS}$ are the strong Fermi spots near M
points (Fig. 1a), labeled as S1 and S2 near the M2(-$\pi$,$\pi$)
point, and S3 and S4 near the M3(-$\pi$,-$\pi$) point. In these two
independent measurements near M2 and M3 (Fig.1a) under different
geometries, the strong spots are only obvious along the $\Gamma$$-$M
direction, with a distance between them (S1 and S2, or S3 and S4) at
$\sim$0.65 $\pi$/a. To check whether this could be affected by the
matrix element effect associated with the particular electric
vector, we took another measurement with a different polarization
geometry  and observed similar pattern (Fig. 1b), suggesting a weak
polarization dependence. These strong spots disappear above the
magnetic transition temperature, T$_{MS}$, (Fig. 1c), and therefore
are intimately related with the magnetic ordering state.

In addition to the strong spots, there are also intensity patches on
both sides of the $\Gamma$-M2(M3) line(s) (Fig. 1a). Band structure
measurements (Fig. 2c and 2f for the Cut $\#$3 in Fig. 1a) indicate
they originate from the electron$-$like bands. The Fermi surface
topology near the M point below T$_{MS}$ is therefore composed of
two electron$-$like Fermi surface sheets and two strong Fermi spots,
as schematically shown in the bottom-right quadrant of Fig. 1a. Note
that the strong Fermi spots are not part of the electron$-$like
Fermi surface sheets because they originate from different bands
(Fig. 2).

It is well$-$known that the conventional SDW formation like in
Cr\cite{CrSDW} is associated with a gap opening along the nested
section of Fermi surface\cite{CrARPES}. It remains under debate
whether similar mechanism operates in the magnetic ordering state in
the parent compound of FeAs-based
superconductors\cite{DongSDW,J1J2Picture}. The detailed Fermi
surface and band structure results allow us to directly examine on
this issue. Considering the nesting vector in BaFe$_2$As$_2$ close
to (0,0)-($\pi$,$\pi$)\cite{DongSDW}, the most likely locations to
search for possible gap opening are close to the Fermi surface
sections near the $\Gamma$-M line (Cut $\#$2 and its equivalent cuts
in Fig. 1a).  It is clear from Fig. 2g that, near the $\Gamma$
point, all photoemission spectra at the Fermi momenta cross the
Fermi level, showing no gap opening along $\Gamma$-M direction.
Measurements along other sections of three Fermi surface sheets near
$\Gamma$ see no signature of gap opening either. Furthermore, there
is no gap opening seen for the strong Fermi spots(Fig. 2h) and for
the two electron$-$like Fermi surface sheets(Fig. 2i). These results
provide a direct evidence that the magnetic ordering state of
BaFe$_2$As$_2$ does not involve gap opening which is fundamentally
different from the conventional SDW system\cite{CrSDW,CrARPES}.

\begin{figure}[tbp]
\begin{center}
\includegraphics[width=0.95\columnwidth,angle=0]{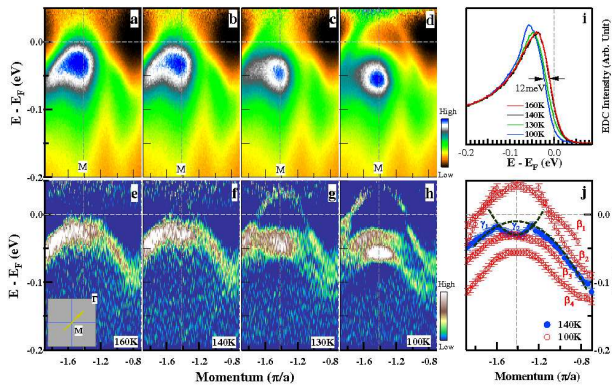}
\end{center}
\caption{Dramatic band reorganization of BaFe$_2$As$_2$ across the
magnetic transition near the M($\pi$,$\pi$) point. (a-d)
Photoemission images taken along the (0,0)$-$($\pi$,$\pi$) direction
near the M3(-$\pi$,-$\pi$) point at different temperatures. The
location of the cut is similar to the Cut $\#$2 in Fig. 1a and also
shown in the inset of Fig. 3e. These images are obtained from the
original data divided by the Fermi distribution function which
allows to observe the electronic states above the Fermi level. (e-h)
The corresponding second derivative images. (i). Photoemission
spectra (EDCs) at the M point measured at different temperatures.
(j) Measured bands at 140 K (blue solid circles) and 100 K (red
empty circles). The data involve symmetrization with respect to the
M point. The black dashed lines are guide to the eyes showing the
band above T$_{MS}$ appears to be composed of two bands: one
hole-like ($\gamma$$_1$) and one electron-like ($\gamma$$_2$).}
\end{figure}

The band structure of BaFe$_2$As$_2$  experiences a dramatic
reorganization across T$_{MS}$=138 K near the M point, in terms of
both the number of bands and their energy position(Fig. 3). Above
T$_{MS}$, only one major branch of band is resolved (Fig. 3a,b,e and
f). As plotted in a more quantitative manner in Fig. 3j, it appears
to be composed of a hole$-$like band ($\gamma$$_1$) and an
electron$-$like band ($\gamma$$_2$).  Below T$_{MS}$, four bands
emerge with inverse$-$parabolic shape (Fig. 3c,d,g,h and Fig. 3j):
two of them ($\beta$$_1$ and $\beta$$_2$ in Fig. 3j) crossing the
Fermi level while the other two below the Fermi level ($\beta$$_3$
and $\beta$$_4$ in Fig. 3j). Such a band structure change is
remarkable considering its occurrence within a narrow temperature
window between 140 K (Fig. 3b and f) and 130 K (Fig. 3c and g). This
also indicates that ARPES signal is sensitive to the magnetic
transition at a temperature that is consistent with the bulk
BaFe$_2$As$_2$ material.

The striking difference of the band structure of BaFe$_2$As$_2$
above T$_{MS}$ (Fig. 4a) and below T$_{MS}$ (Fig. 4b) clearly
indicates that the magnetic ordering state does not inherit its
electronic structure simply from its high-temperature non-magnetic
counterpart. A natural question is whether such a strong band
reorganization can be attributed to the band-folding effect
associated with the magnetic$\slash$structural transition. Such a
transition renders the initial $\Gamma$ and M points equivalent in
the new shrunk Brillouin zone causing the bands at $\Gamma$ and M to
fold to each other. In principle, such a band folding effect is
always present regardless of the origin of the magnetic ordering. In
real experiment, whether the folded bands can be resolved depends on
the interaction potential and matrix element effect: the intensity
of the folded bands can be much weaker than the primary
ones\cite{Voit,Brouet}. Such a folding effect was previously
proposed to account for the magnetic transition in
SrFe$_2$As$_2$\cite{ZHasanSFS}. But our observations indicate it is
not the main contributor to the dramatic band reorganization (Fig.
4). The two hole-like bands below the Fermi level near M
($\beta$$_{b3}$ and $\beta$$_{b4}$ in Fig. 4b) definitely can not be
due to the band folding effect because there are no corresponding
bands near $\Gamma$. For the same reason, the inner-most band near
$\Gamma$ ($\alpha$$_{b3}$ in Fig. 4b) can not be due to such a band
folding from M either. As for the two hole-like bands crossing the
Fermi level near M ($\beta$$_{b1}$ and $\beta$$_{b2}$ in Fig. 4b),
they appear similar to the bands near $\Gamma$ and one may wonder
whether they may originate from the folding effect.  However, from
the bands in Fig. 3j where the bands near and above E$_F$ are
clearly visible (Fig. 3), the two hole-like bands cross each other
near E$_F$. This observation makes them unlikely to originate from
the two bands near $\Gamma$ which are separate.  The band folding is
also expected to produce similar Fermi surface topology near
$\Gamma$ and M. While the strong Fermi spots are obvious near M, no
such replica can be identified near $\Gamma$ point(Fig. 1a). This
further indicates that the folding effect in BaFe$_2$As$_2$ is weak
although in principle it is present.

The strong Fermi spots are most prominent electronic signature of
magnetic ordering in BaFe$_2$As$_2$ as they only appear below
T$_{MS}$ (Fig. 1a and c). One natural way to form such a singular
Fermi spots is to have two bands crossing near the Fermi level
giving rise to a Dirac cone-like structure. In FeAs-based compounds
with multiple bands, below T$_{MS}$, the folding between the
hole-like bands near $\Gamma$ and the electron-like bands near M can
give rise to such Dirac nodes along $\Gamma$-M direction which are
gapless\cite{YRan}. These characteristics seem to be in qualitative
agreement with our observations and similar folding picture was also
used for a previous ARPES experiment\cite{ZHasanSFS}. However, some
of our experimental observations are at variance with this picture.
In the folding picture, the Dirac nodes are produced from one
hole-like band folded from $\Gamma$ to M and one electron-like band
near M\cite{YRan,ZHasanSFS}.  However, our results indicate that the
spectral weight of the strong spots are mainly from the hole-like
bands (Fig. 2b and 2e).  Fig. 3j further suggests that the strong
spots are more likely formed by the two hole-like bands
($\beta$$_{1}$ and $\beta$$_{2}$) crossing each other near the Fermi
level.  The contribution to the strong spots formation from the
electron-like bands is weak because no trace of electron-like bands
is resolved  along the $\Gamma$-M direction(Figs. 2b, 2e and Fig. 3)
probably due to matrix element effect.

We note that the strong Fermi spots we have observed (Fig. 1a) show
some resemblance to the ``propeller-shaped" pattern reported before
for a given polarization in both BaFe$_2$As$_2$ and
(Ba$_{1-x}$K$_{x}$)Fe$_2$As$_2$ superconductor which was attributed
to ``hidden" ($\pi$,$\pi$) excitations that exist all the way to
room temperature\cite{Borisenko}. Our results clearly indicate that
the strong Fermi spots can not originate from the same ``hidden"
($\pi$,$\pi$) order because they appear only below T$_{MS}$ (Fig. 1
and Fig. 3).  We also observed strong Fermi spots near M in
(Ba$_{0.6}$K$_{0.4}$)Fe$_2$As$_2$ superconductor\cite{LZhao}. In
terms of the Fermi surface topology, our observation of strong Fermi
spots in both magnetically$-$ordered BaFe$_2$As$_2$ and
superconducting (Ba$_{0.6}$K$_{0.4}$)Fe$_2$As$_2$ seems to be
consistent with the previous experiment showing similar
patterns\cite{Borisenko}. However, the underlying band structure we
have revealed between them is remarkably different\cite{LZhao}. This
indicates that the seemingly similar strong Fermi spots in the
parent (Fig. 1) and K-doped BaFe$_2$As$_2$\cite{LZhao} are most
likely from different origins, rather than from the same
($\pi$,$\pi$) excitations proposed before\cite{Borisenko}.

\begin{figure}[tbp]
\begin{center}
\includegraphics[width=0.86\columnwidth,angle=0]{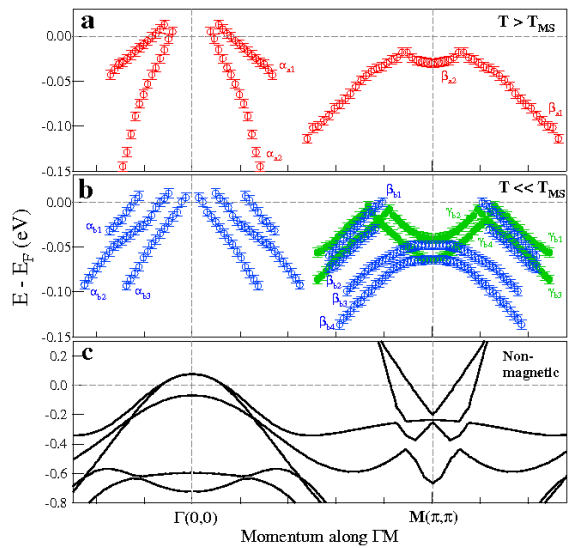}
\end{center}
\caption{Band structure of BaFe$_2$As$_2$ along the $\Gamma$$-$M
high symmetry line above (a) and below (b) T$_{MS}$. Below T$_{MS}$
(Fig. 4b), the bands near the M point are from two measurements: Cut
$\#$2 in Fig. 2b (blue empty circles) and Cut $\#$3 in Fig. 2c
(green solid circles). (c). Electronic structure of BaFe$_2$As$_2$
from band calculations in a non-magnetic
state\cite{Nekrasov,DJSinghBFA,FJMa}.}
\end{figure}

The dramatic band reorganization across T$_{MS}$ and the lack of gap
opening provide clear-cut evidences that the magnetic ordering state
of BaFe$_2$As$_2$ is distinct from the usual SDW
state\cite{CrSDW,CrARPES}. These observations may provide a new
approach to understand previous
measurements\cite{Oscillations,WZHuOpt}. For example, the multiple
bands developed below T$_{MS}$ (Fig. 4b) may provide new channels
for inter-band transitions, offering an alternative interpretation
of gap opening in terms of SDW picture that was proposed in the
optical spectroscopy\cite{WZHuOpt}. Our results show that the
multiple-band structure (Fig. 4b) and the appearance of strong Fermi
spots (Fig. 1a) below T$_{MS}$ are neither inherited from its high
temperature non-magnetic state, nor from the band folding effect
associated with the magnetic$\slash$structural transition. These
indicate that they are electronic fingerprints inherent to the
magnetic ordered state. These characteristics reflect the electronic
complexity in this multiple-orbital system and point to the
importance and necessity in involving the local magnetic moment in
the underlying electron structure. How to describe the peculiar
electronic features of the magnetically ordered state on a
quantitative level presents a challenge to further theoretical work.
Our observations and their understanding will shed new light on
understanding the magnetism and high temperature superconductivity
in FeAs-based superconductors.

We thank A. V. Balatsky, J. P. Hu, D. H. Lee, Z. Y. Lu, J. L. Luo,
E. W. Plummer, N. L. Wang and T. Xiang for helpful discussions. This
work is supported by the NSFC, the MOST of China (973 project No:
2006CB601002, 2006CB921302).

$^{*}$Corresponding author: XJZhou@aphy.iphy.ac.cn


\end{document}